\documentclass[12pt,preprint]{aastex}

\slugcomment{Accepted to ApJ}

\shorttitle{Cataclysmic and Close Binaries in M80}
\shortauthors{Shara et al.}

\begin{document}

\title{Cataclysmic and Close Binaries in Star Clusters. V. Erupting 
Dwarf Novae, Faint Blue Stars, X-ray sources, and the Classical Nova in the Core of M80\footnotemark[1]}

\footnotetext[1]{Based on observations with
 the NASA/ESA {\it Hubble Space Telescope},
 obtained from the Data Archive at the Space Telescope Science Institute,which is 
operated by the Association of Universities for Research in Astronomy, Inc., 
under NASA contract NAS5-26555. These observations are associated with program \#6460.}

\author{Michael M. Shara}
\affil{Department of Astrophysics, American Museum of Natural History, 79th St. and Central Park West, 
New York, NY, 10024}
\email{mshara@amnh.org}

\author{Sasha Hinkley}
\affil{Department of Astrophysics, American Museum of Natural History, 79th St. and Central Park West, 
New York, NY, 10024}
\email{shinkley@amnh.org}

\author{David R. Zurek}
\affil{Department of Astrophysics, American Museum of Natural History, 79th St. and Central Park West, 
New York, NY, 10024}
\email{dzurek@amnh.org }


\begin{abstract}
Large populations of cataclysmic variables (CVs) in globular clusters 
have long been predicted, but the number of absolutely certain cluster CVs known in 
globulars is still less than 10. HST and Chandra observers have recently 
found dozens of cataclysmic variable candidates in several populous 
globular clusters. Confirmation and characterization of these candidates 
are extremely difficult, thus identification of unambiguous CVs remains 
important. We have searched all archival HST images of the dense 
globular cluster M80 for erupting dwarf novae (DN), and to check the outburst 
behaviors of two very blue objects first identified a decade ago. Two 
new erupting dwarf novae were found in 8 searched epochs, making M80 a 
record holder for erupting DN. The quiescent classical nova in 
M80 varies by no more than a few tenths of a magnitude on timescales of minutes to 
years, and a similar faint, blue object varies by a similar amount. 
Simulations and completeness tests indicate that there are at most 3 erupting DN 
like SS Cyg and at most 9 U Gem-like DN in M80. Either this very dense cluster contains 
about an order of magnitude fewer CVs than theory predicts, or most M80 CVs are extremely faint 
and/or erupt very infrequently like WZ Sge. We have detected a sequence of 54 
objects running parallel to the main sequence and several tenths of a magnitude 
bluewards of it. These blue objects are significantly more centrally concentrated
than the main sequence stars, but not as centrally concentrated 
as the blue stragglers. We suggest that these objects are white dwarf--red dwarf binaries, and that some are the faint CV 
population of M80.
\end{abstract}


\keywords{Stars: Cataclysmic Variables--- dwarf novae, galaxies: 
individual (LMC)}


\section{Introduction}

The discovery of unexpectedly rich populations of luminous X-ray sources 
in Galactic globular clusters \citep{cla75, kat75} was a sharp 
challenge to theorists. Efficient mechanisms for generating large 
numbers of mass-transferring binaries, dominated by a compact primary, 
were required to explain the observations. Tidal capture scenarios 
involving two stars \citep{fab75} and/or exchange reactions involving 
three stars \citep{hil76} were rapidly developed. These processes are 
now believed to be the sources of the neutron star-dominated X-ray 
binaries in globular clusters. Strong observational evidence in favor of 
these scenarios is the remarkable correlation between stellar encounter 
rate and number of X-ray sources in globular cluster cores \citep{poo03}.

These two and three-body mechanisms also predict large populations 
\citep{dis94} of accreting white dwarf-main sequence star binaries---the 
cataclysmic variable (CV) stars. Unfortunately only one globular cluster 
dwarf nova (DN) is easily resolved and studied from the ground: V101 in 
M5 \citep{mar81}, \citep{sha90}, \citep{nei02}. The large amplitude 
variations of many CVs \citep{war95} suggest that Hubble Space Telescope 
(HST) observations of globular clusters would easily locate dozens in the core 
of each cluster. With expected apparent magnitudes in the range 17-23, 
prototypical CVs like SS Cyg and U Gem should be trivial to find. It 
would clearly be of enormous benefit to CV science if dozens of these 
objects, all at the same distance and with the same parent metallicity, 
could be located. Accurate luminosity functions, and bias-free period 
distributions and eruption frequencies could be derived to confront 
theoretical models. Tidal capture CVs could be contrasted with those 
produced by ordinary binary evolution. Systematic variations in outburst 
properties and orbital distributions might be uncovered.

Remarkably, only a handful of absolutely certain CVs in globular 
clusters are known today. These include a few dwarf novae seen in 
eruption (e.g. \citep{par94}, \citep{cha02}, \citep{and03}), one 
classical nova \citep{auw86}, \citep{sha95}, and a few spectrographically 
confirmed, very blue faint stars
(e.g. \citep{gri95}, \citep{deu99,kni03}). Recently, dozens of CV 
candidates have been identified in 47 Tuc \citep{ghe01}, 
\citep{kni02}, M80 \citep{hei03} and in NGC 6397 \citep{gri01} through 
deep Chandra X-ray Observatory and HST observations. Unfortunately, large 
amounts of HST and Chandra time are essential for follow-up studies, so 
unambiguous characterization of all candidates may take many years. Thus any new, 
unambiguous CVs that can be found in archival data are well worth 
searching for.

About half of all known field CVs \citep{dow01} are dwarf novae (DN). 
Most known DN reveal themselves through 2-5 magnitude outbursts every 
few weeks to months \citep{war95}; similar DN in clusters should be 
identifiable in multi-epoch images. Since large archival HST datasets of 
globulars are available we have been systematically looking for erupting 
dwarf novae in the cores of all such clusters. Here we report the results 
of an extensive and successful search of M80.

M80 is one of the densest globular clusters in the Milky Way, with a 
spectacular blue straggler sequence \citep{fer99}, a quiescent old nova 
\citep{sha95}, and 19 X-ray sources \citep{hei03}. A detailed comparison 
of M80 with 47 Tuc \citep{hei03} suggests that the encounter rate in M80 
is about a factor of two lower; roughly 50 CVs should exist 
in M80 as of order 100 are predicted in 47 Tuc \citep{dis94}.

In section 2 we present the observational database, search strategy, and 
the photometry of variables. The erupting DN, non-eruptive CV candidates 
and their light curves are shown in section 3. We discuss whether the 
candidates could be other kinds of variables in section 4. In section 5 
we report simulations to test our completeness, and to place limits on 
the numbers of DN of various luminosities in M80. We report on a remarkable 
sequence of very blue, centrally located objects in section 6, and our attempts
to match these with the M80 X-ray sources in section 7. We briefly summarize 
our results in section 7. 

\section{Observations}
The Hubble Space Telescope has imaged M80 during 8 separate epochs from 
1994 to 2000. All epochs except the last included F336W frames, 
particularly useful in detecting blue objects. The pass-band of the 
F450W frames of the last epoch is close to the F439W passband of epochs 1 
and 2, so that essentially constant limiting magnitude is attained for 
our 8 epoch survey (see section 5). The dates of observation, PI of program
and program number, filters used, number of frames and total exposure time in each filter 
are given in the observing log which is Table 1. 

The individual WFPC2 images were combined using the ``Montage2'' 
routine contained within the stand-alone DAOphot package. 
These frames were run through DAOphot's matching program ``DAOmaster'' 
to derive the subpixel frame-to-frame shifts and thereby register all 
images to the first image of the first epoch. These shifts were passed to 
Montage2, which produced a single, sky subtracted, high signal-to-noise 
image for each epoch and filter. HST was not perfectly aligned at all 
epochs in all filters. Figure 1 is an overview of the fields of view of 
the 8 epochs, with four of the most interesting objects that we found 
(section 3) indicated.

The dwarf-nova candidates (Figure 2) were found by image differencing {\it and} by blinking rapidly through 
groups of the F336W frames to look for any subtle changes in the 
appearance of the stellar field. Dwarf novae should rise from invisibility to 
easily detectable in the timeframe defined by the observations. Brightening from 
previously empty regions of the sky is easily seen. 

In order to detect extremely faint stars, a master list of stars was derived
using the technique of \citet{ric04}. This technique rejects the brigher pixels from
a stack, giving emphasis to fainter pixels, and thus probing deeper. In addition, point spread 
functions were created for the F336W, F450W, F656N, and F675W ($U$, $B$, $H\alpha$ and $R$) data and then 
provided to the allframe software \citep{ste94} for multi-image simultaneous PSF fitting to derive 
positions and magnitudes for as many stars as possible. The photometry was then 
converted to the ST magnitude system as described in \citet{sha95}. The resulting 
color-magnitude diagram (CMD) of M80, the deepest ever produced, is shown in Figures 3 and 4. 
The two new dwarf nova candidates are evident, as are the two very blue stars first noted 
by \citet{sha95}. In addition, many more faint blue objects---about 50 in all---are now 
visible in this deeper CMD. Some of these very blue objects are also indicated in the finder chart 
that is Figure 5, and discussed in sections 6 and 7. 

\section{Dwarf Novae and the Classical Nova}

To be recognized as a candidate, a variable had to be visible on all frames in at 
least one of the eight epochs, and to have varied by more than 1.0 magnitudes 
between its faintest and brightest state. This selection method revealed 
two dwarf nova candidates from the entire dataset (and no other obvious variables).  
These two strongly varying, very blue 
candidates which we label DN1 and DN2 are erupting in epochs 3 and 6, respectively. 
Their positions are: for DN1 $\alpha(2000)=$ 16:17:02.2, $\delta(2000)=$ -22:58:37.9; 
and for DN2, $\alpha(2000)=$16:16:59.8, $\delta(2000)=$ -22:58:18.0. 
The nightly images of these two objects are shown in Figures 6 and 7, 
and their photometry is presented in Figures 8 and 9. We defer a discussion of these 
stars and their implications until after the next section, where we simulate expected 
DN images and light curves.

We show the images and photometry of the likely quiescent old nova (T Sco 1860 A.D.) 
in Figures 10 and 11, respectively. Brightness fluctuations do not exceed a few tenths 
of a magnitude. The mean apparent F336W 
magnitude is 21.5, corresponding to an absolute U-band magnitude of 6.5, well within 
the range of old nova luminosities \citep{war95}.

Finally, we show in Figures 12 and 13 the images and lightcurve of the faint blue 
object noted by \cite{sha95}. Located close to T Sco in the M80 color-magnitude 
diagram, it too varies by at most a few tenths of a magnitude. Whereas the T Sco 
lightcurve shows several points in a given epoch, the faint blue object's lightcurve
shows only one brightness value for each epoch. The object lies rather close to the 
HST Planetary Camera (PC) chip's edge, and photometry was performed on the 
individual-epoch, CR-rejected images to boost the signal-to-noise. While its brightness 
and color are very suggestive, this object remains a CV candidate, and not a confirmed CV, 
until it is eventually shown to erupt as a DN or found to have a spectrum consistent 
with cataclysmic classification.  

\section{Simulations of Known Dwarf Novae}

To estimate our detection completeness for erupting dwarf novae to various magnitude limits 
we carried out simulations 
of eruptions for two of the best known and characterized Galactic 
dwarf novae: SS Cyg, and U Gem. (These objects have accurately determined 
parallaxes \citep{har99} and, using their well tabulated apparent magnitudes 
\citep{war95}, the absolute magnitudes are easily calculated. These 
two dwarf novae were chosen to represent, respectively, the most luminous and more typical
values of absolute magnitudes seen for erupting and quiescent DN). To simulate SS Cyg, 
100 artificial PSFs were placed in random positions across an M80 HST F336W image of a single 
epoch. For these 100 PSFs, brightnesses were determined by picking 100 random points from 
the SS Cyg American Association of Variable Star Observers (AAVSO) lightcurve stretching over 
1000 days. These 100 artificial dwarf novae, (now mimicking SS Cyg at 100 random places in 
its lightcurve), were shifted to the apparent magnitudes they would have at the 10.3 kpc distance 
of M80 \citep{bro98}. This image was then blinked against an image (at a different epoch) with 
these same artificial PSFs corresponding to SS Cyg in quiescence. The same procedure was 
performed for U Gem, using 100 randomly sampled brightnesses from 1000 days of the U Gem lightcurve.
Typical images of these simulated dwarf novae, each at their maximum, intermediate and 
minimum brightness, are shown in Figure 14. Our simulations show that, by comparing two randomly selected 
epochs, one detects an SS Cygni-like dwarf nova $25.0 \pm 3.3 \%$ of the time, while U Gem is
detected $6.0 \pm 1.7\%$ of the time. The detection probability is thus $P_{SS}=0.125$ per epoch for 
SS Cyg and $P_{UG}=0.030$ for U Gem. In each independent epoch, we therefore expect that the probability
of {\it not seeing} a given SS Cyg in eruption is $1 - 0.125 = 0.875$, and the probability of 
not seeing a given U Gem in eruption is $1 - 0.030=0.970$. Thus in eight independent epochs (a more 
pessimistic case than our own 8 epochs, where 5 epochs are precisely at weekly intervals) we might
expect to not see SS Cyg $(0.875)^8=34\%$ of the time, and we expect to miss U Gem-like objects 
$(0.97)^8=78\%$ of the time. Equivalently, we expect to see at least $1-0.34 = 66\%$ of SS Cyg-like
eruptions and at least $1-0.78=22\%$ of all U Gem-like eruptions.

\section{Interpretation}
	
\subsection{Have We Found Erupting M80 Dwarf Novae?}

Comparison of the two erupting DN candidates shown in Figures 6 and 7 
with the M80-DN simulation images shown in Figure 14 demonstrates that the 
brightness and variability behaviors of our candidates are consistent with those expected of 
moderately luminous erupting dwarf novae in M80. All erupting dwarf novae achieve outburst U band 
absolute magnitudes $\lesssim 5$, and the simulations of Figure 15 show that we easily reached 
fainter than that limit in all epochs. The very blue outburst colors, outburst 
brightnesses, presence and blue colors in quiescence, our completeness of 
detection of stars at different magnitudes (Figure 15), and strong 
concentration to the center of M80 all argue very strongly in favor of 
DN1 and DN2 being dwarf novae. However, in the absence of spectra or a 
second recorded outburst, one could always argue that one or both 
of these are variables of some other type, perhaps not even associated with M80.

What are possible variables that might mimic M80 DN behavior? Amongst 
these are: chance superpositions of background supernovae or classical 
novae, gamma ray bursts (GRB), microlensing events or Milky Way 
variables along the line of sight to M80. Large area, multi-epoch 
surveys for faint variables (e.g. \citet{haw84}) 
show that the strong central concentration of our variables to the core 
of M80 (and the moderately high galactic latitude of the cluster, $19.5^{\circ}$) is far too high for them to be field RR Lyrae stars or 
supernovae.

The brightness and very blue colors of the candidates rule out other types of 
Galactic variables. RR Lyrae and flare stars don't match the 
observed brightness and/or blue colors of the two DN candidates. The 
rarity of GRB (about 1 over the entire sky per day) and microlensing 
events, and the very blue colors of our variables almost certainly rules 
out these possibilities. But while the magnitudes and colors both during 
outburst and quiescence of our DN candidates argue strongly for that 
classification, an absolutely certain characterization will require 
challenging but important follow-up observations. These include

1) Spectra near quiescence (to demonstrate the presence of Balmer 
emission lines).

2) Imagery every day or two for several months with HST to reveal 
repeated eruptions separated by weeks to months.

3) Several hours of HST time-resolved UV or optical photometry to reveal 
the orbital modulation characteristic of CVs.


\subsection{Detections versus Expected Detections}

The size of our field of view was 1574 arcsec$^2$, extending 
out to 1.9 core radii from the cluster center. The 
average time between eruptions for 21 well studied Galactic DN is 29 days 
\citep{szk84}. The length and depth of our observing run (5 epochs spaced 1 week 
apart, and three other random epochs) taken by themselves suggest that most 
erupting SS Cyg-like or U Gem-like dwarf nova would have been detected. 
In fact, our simulations of SS Cyg and especially of U Gem (in section 4) 
demonstrate that outburst frequency is the key parameter in determining whether
a given dwarf nova will be detected. SS Cyg is intrinsically more luminous in 
outburst than U Gem, but both are straightforward to detect, in eruption, in 
our dataset. It is the relative infrequency of U Gem outbursts that makes it three 
times less likely (22\%) to be detected than SS Cyg-like outbursts. Given these probabilities
of detecting SS Cyg and U Gem-like dwarf nova eruptions in M80, what limits can we place 
on the total populations of similar objects in the cluster? Since we have only seen 1 eruption 
of each of the M80 DN we don't know their eruption frequency or maximum luminosity. Either could
be U Gem-like or SS Cyg-like. Our conservative estimate is that there are $\lesssim 2/0.66 = 3$ SS Cyg-like
DN in M80, and $\lesssim 2/0.22 \sim 9$ U Gem-like DN in the cluster. 

The fact that we detected only two erupting DN, and that there are almost certainly fewer than $\sim9$
DN in M80, when (the admittedly 
simple) tidal capture model predicts an order of magnitude more CVs suggests that

i) SS Cyg and U Gem-like CVs in M80 are $\sim50/9\sim6\times$
fewer in number than theory predicts, and/or

ii) most CVs are more like the rarely erupting WZ Sge, and much fainter than the 
prototypical DN SS Cyg and U Gem, and/or

iii) globular CVs may be mostly magnetic \citep{gri95}and non-eruptive. 

What is the true, average inter-outburst period for all dwarf novae? Unfortunately, 
the answer is unknown. CV catalogs are dominated by the 
easy-to-find, frequently outbursting DN. In addition, the distances to all but a dozen 
or so Galactic DN are too uncertain to yield an accurate luminosity 
distribution. (We note that the canonical literature distance to the 
prototypical DN, SS Cyg, was in error by a factor of two until HST 
parallaxes became available in 2000). A warning that CV populations may 
be dominated by extremely faint and/or infrequently erupting DN is 
personified in the nearest known DN, WZ Sge. This object is a mere 43.5 
pc away \citep{tho03}, \citep{har04} with absolute visual magnitudes at 
minimum and maximum of 11.8 and 3.9, and 1--2 month-long eruptions separated 
by about 25 years. Our chances of detecting a WZ Sge-like outburst are two 
orders of magnitude smaller than the likelihood of detecting a U Gem-like 
eruption (again, because of the relative outburst frequency). There could easily 
be 100 WZ Sge stars in M80---and our chances of detecting even one in our dataset 
would only be of order 20\%. We conclude that: there are $\sim3$ or fewer SS Cyg-like,
or 9 or fewer U Gem-like DN in M80. However, 100 WZ Sges could exist in the cluster. 
If most CVs are similar to WZ Sge (often referred to as CV 
``graveyard'' objects) then the small number of detected M80 DN is 
naturally explained.

\section{The Faint Blue Objects in M80}

The deep $U$ vs. $(U - B)$ CMD of Figure 5 contains a remarkable sequence of 54 
objects running parallel to the main sequence and between 0.3 and 1.4 magnitudes blueward 
of it. We have verified that these faint blue objects are not due to e.g. ``hot'' pixels by comparing 
their images on frames from different epochs with significant spatial offsets. 
While it is certainly possible that some of these objects' blue colors are due to 
photometric errors, at least a dozen objects with $(U - B) < 0.1$ are (on visual 
inspection) very blue and unblended. This is precisely the part of the CMD that 
would be occupied by CVs and non-interacting white dwarf-red dwarf binaries. If these 54 objects 
are, indeed CVs and/or WD-RD binaries then they must possess masses larger than main sequence 
stars of similar magnitude. Equipartition of energy should then have concentrated these 
blue objects towards the cluster center. This, in fact, is an acid test that must be 
passed if we are to give credence to the suggestion of the blue M80 stars as a WD-RD 
binary sequence. Figure 16 strongly suggests that these blue stars are more centrally 
concentrated than main sequence stars, at least at $r > 8$ arcsec from the cluster center. 
The K-S test determines that the blue objects and main sequence stars of Figure 16 do not 
share the same radial distribution with $ > 95 \%$ confidence. The blue stragglers and main
sequence stars are even more dramatically separated in Figure 16. The K-S test states that the 
BS and MS stars are drawn from different radial distributions with $> 99.5\%$ likelihood.

Artificial star tests were performed to determine the 
the recovery rate for artificial stars placed randomly on the chip. Figure 17 shows the fraction of 
stars recovered in 8 different annuli centered on the center of the cluster, for several different 
brightesses. These simulations show that we approach maximum completeness ($\sim80\%$)
for the fainter blue stars with $U > 
22$ only when $r > 8$ arcsec from the core of M80. {\it Thus the apparent weak central concentration of 
the blue stars in the inner 8 arcsec is probably due to significant incompleteness of the sample 
of these very faint objects in the core of the cluster, rather than their physical absence.}  

The central concentration of blue stragglers in globular clusters is ubiquitous and well 
established for many clusters e.g. \citep{par91,fer99}, and indicative of objects with measured masses 
2-3 times the turnoff mass \citep{sha97,saf02}. We conclude from Figure 16 that the blue sequence 
stars of Figure 13 are intermediate in mass between the main sequence and blue straggler stars 
---exactly as one expects for WD-RD binaries.

\section{X-ray sources and Faint Blue Objects}
We have compared the positions of our four objects of interest (DN1, DN2, T Sco and the Shara-Drissen 
object) as well as the 54 faint, blue 
objects with the 19 X-ray sources noted by \citet{hei03}. The Right Ascension and Declinations 
of our full sample of 58 CVs and CV candidates were offset in an iterative manner until we found 
the optimal minimum separation between the 19 X-ray sources and the CV candidates. The best overall match 
(in the least-squares sense) occurs with offsets of $\Delta\alpha\simeq -0.5''$ and $\Delta\delta\simeq +4.7''$ 
applied to the CV candidates, with or without the two significant outliers CX10 and CX19 removed. 
The resulting source distributions are shown in Figure 18. The 19 X-ray source positions together 
with the adjusted positions of the closest faint blue stars are listed in Table 2. 

The nearly 5'' offset between the 19 X-ray sources and the 58 CVs and CV candidates is much larger than the 
uncertainties in HST and Chandra pointings. We also find no correlation between the 19 X-ray sources' luminosities 
and the F336W brightnesses of the 19 ``matched'' faint blue stars in Table 2. We interpret this to mean that, 
while the two populations are centrally concentrated, many (and probably most or all) of the apparent matches of Table 2 are 
spurious. Deep HST far-UV observations of M80 are currently being scheduled, and these will be important for 
providing optical-UV counterparts to this cluster's X-ray sources. 

Finally, we have also checked the H$\alpha$ luminosities of the 54 faint blue 
sources. An $R$ versus H$\alpha - R$ color-magnitude diagram is shown in Figure 19. Of the full 
sample of 54 blue objects, 51 had measurable H$\alpha$ fluxes. None of these faint blue 
sources appear to have an H$\alpha$ brightness more significant than that of the overall cluster 
population. If these are degenerate star--red dwarf binaries then their mass transfer rates are very low 
or zero.

\section{Conclusions}

We have observed M80 during 8 separate epochs with sufficient resolution and
sensitivity to detect any erupting dwarf novae within the inner 1.9 core 
radii of that cluster. Two very strong candidates, with outburst colors, 
brightnesses, central concentration, and presence in quiescence, all consistent with DN 
classification, were 
detected. The old nova candidate T Sco (nova 1860) and another very blue, faint star \citep{sha95} are 
remarkably constant in brightness. Simulations and completeness tests indicate that 
$\le3$ regularly outbursting dwarf novae similar to SS Cyg, and $\lesssim 9$ U Gem-like dwarf novae
exist today in M80. The presence of several dozen very faint blue, centrally concentrated stars 
in the core of M80 hints at a large red dwarf--degenerate star binary population; some of these
may be a very rarely-erupting CV population.


\acknowledgments
We acknowledge with thanks the variable star observations from the AAVSO 
International Database contributed by observers worldwide and used in this research.

Support for program \#6460 was provided by NASA through a grant from the Space 
Telescope Science Institute, which is operated by the Association of Universities for 
Research in Astronomy, Inc., under NASA contract NAS 5-26555."

\appendix 


\clearpage


\begin{figure}
\figurenum{1}
\plotone{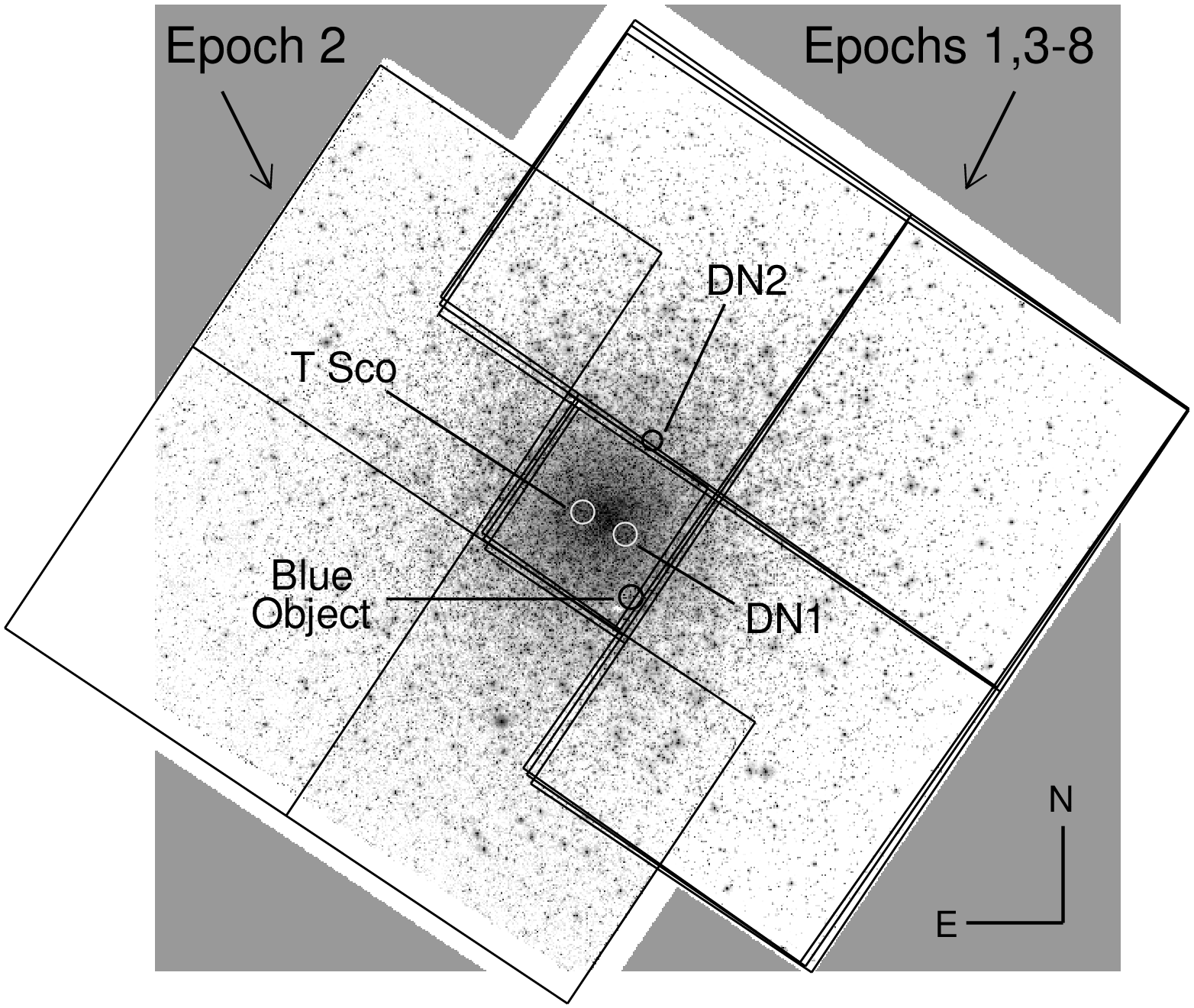}
\caption{A mosaic of HST images of the field of M80, indicating the orientations of our 8 epochs of observation. }
\end{figure}

\begin{figure}
\figurenum{2}
\plotone{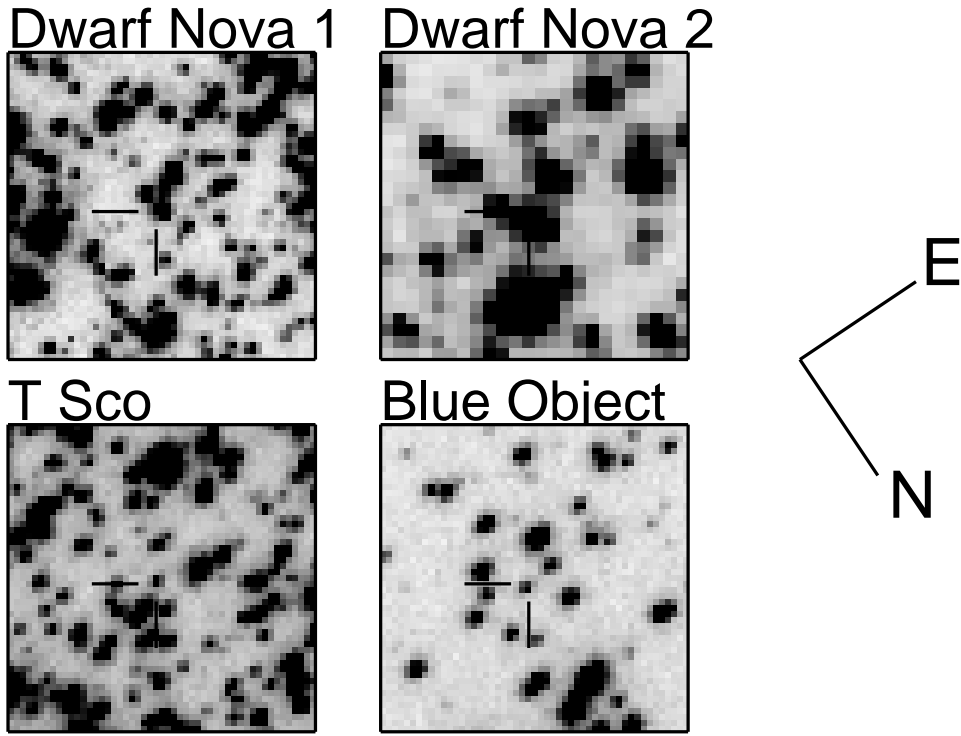}
\caption{A closer view of each of the four objects of interest. Each image is 2.37 arcseconds 
on a side. The lower resolution in the image of Dwarf Nova 2 reflects the fact that this object 
was only imaged on the HST WF4 chip, while all the others were imaged on the higher resolution 
Planetary Camera CCD.}
\end{figure}

\begin{figure}
\figurenum{3}
\plotone{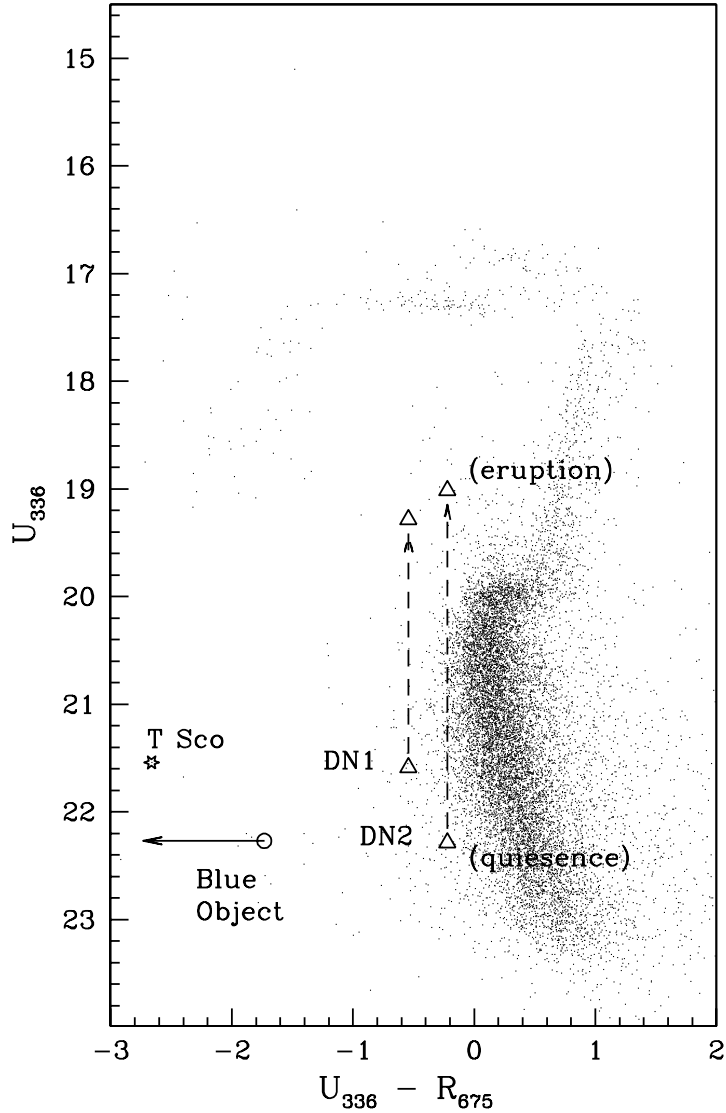}
\caption{CMD for M80. In the plot, the four objects of Figure 1 are marked. Since there was only $R$ imaging for one 
epoch (epoch 3), obtaining both eruption and quiescent {\it colors} was impossible. The range of $U$ 
magnitudes for the two dwarf novae are shown to give a sense of the overall range during outburst. In addition, 
since the Blue Object was essentially invisible in the $R$-band, an $R$ value near the 
plate limit was assumed, hence the upper limit arrow indicated. }  
\end{figure}

\begin{figure}
\figurenum{4}
\plotone{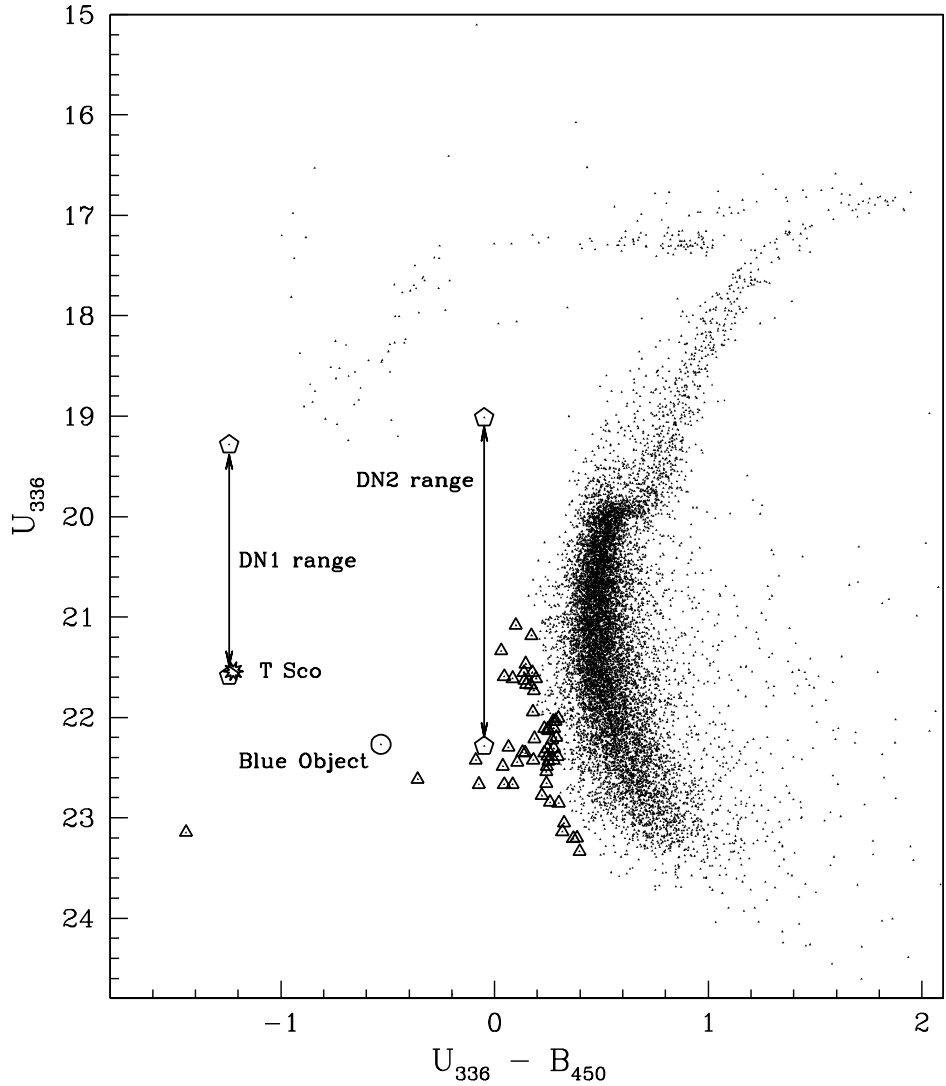}
\caption{ The $U$ versus $(U - B)$ CMD for M80. The 54 triangles that lie blueward of the main sequence 
are newly detected white dwarf-red dwarf binary and/or
CV candidates. The four objects discussed in this paper are also marked. The faint blue object 
of \citet{sha95} is 
marked with a circle, while T Sco (nova 1860) is starred. Note that the location of T Sco is nearly coincident with the first 
dwarf nova (DN1) in its quiescent state. As before, the brightness ranges of the two dwarf novae DN1 and DN2 are marked. }
\end{figure}

\begin{figure}
\figurenum{5}
\plotone{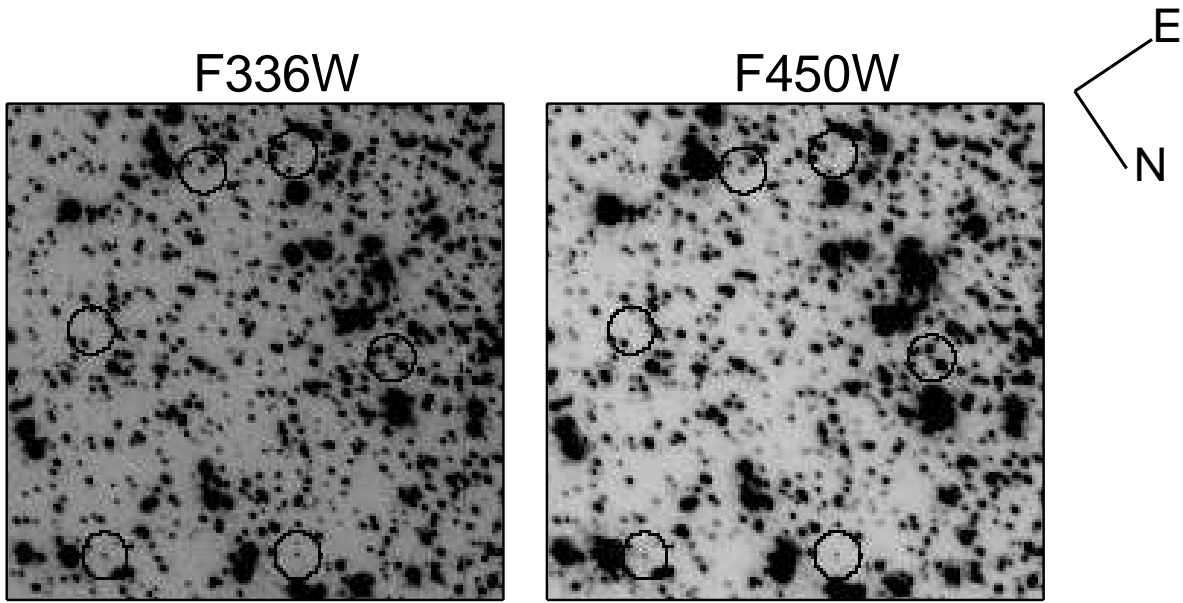}
\caption{Two subregions of the F336W and F450W PC images showing a handful of those objects significantly
more blue than the main sequence. North and East are as indicated. Each image is 8.38 arcseconds on a side.}
\end{figure}

\begin{figure}
\figurenum{6}
\plotone{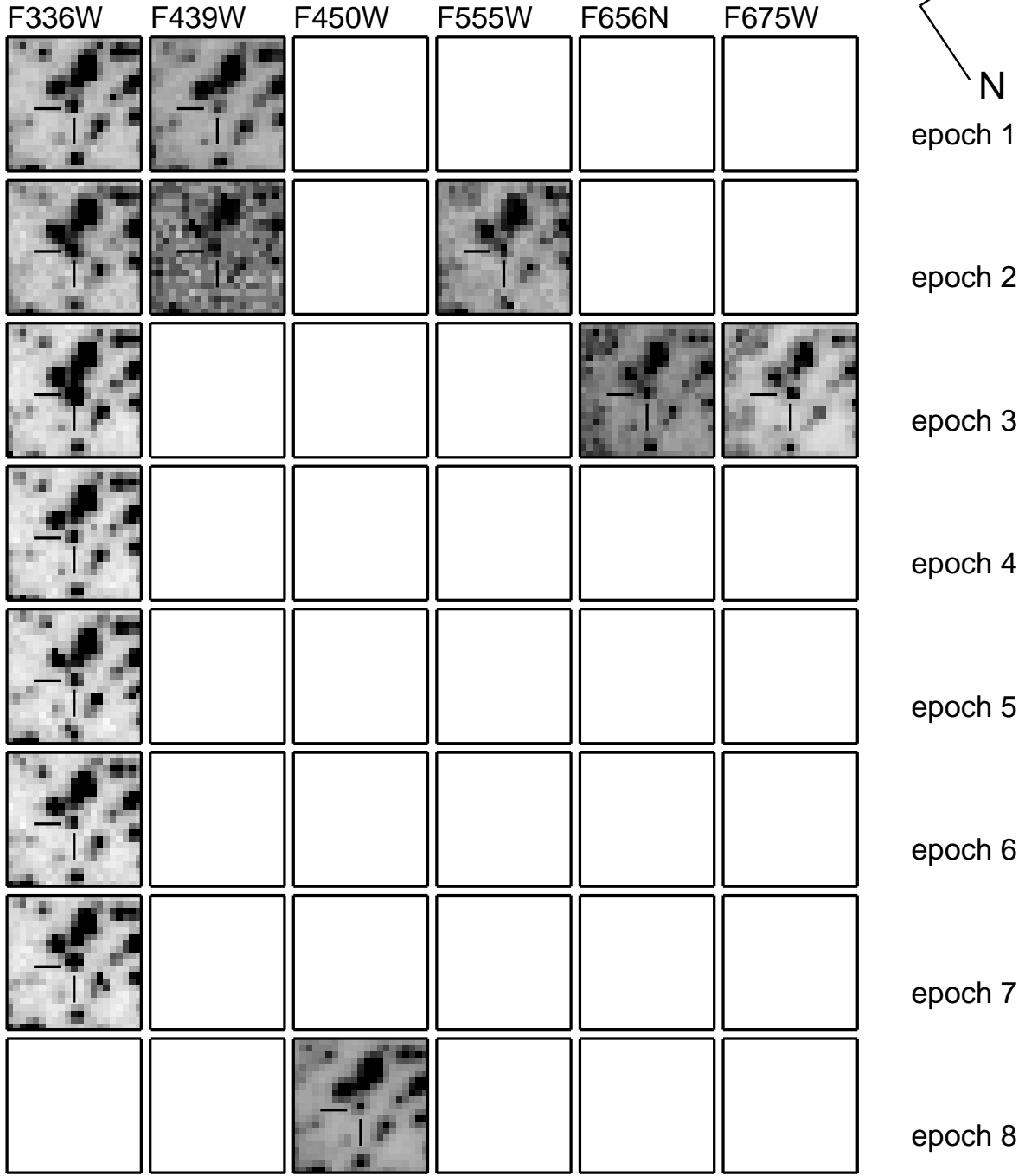}
\caption{Thumbnail images for the first Dwarf Nova (DN1) in M80. It can clearly be seen in eruption in 
epoch 3. Each thumbnail is 1.04 arcseconds on a side, and North and East are indicated at the upper right. }
\end{figure}

\begin{figure}
\figurenum{7}
\plotone{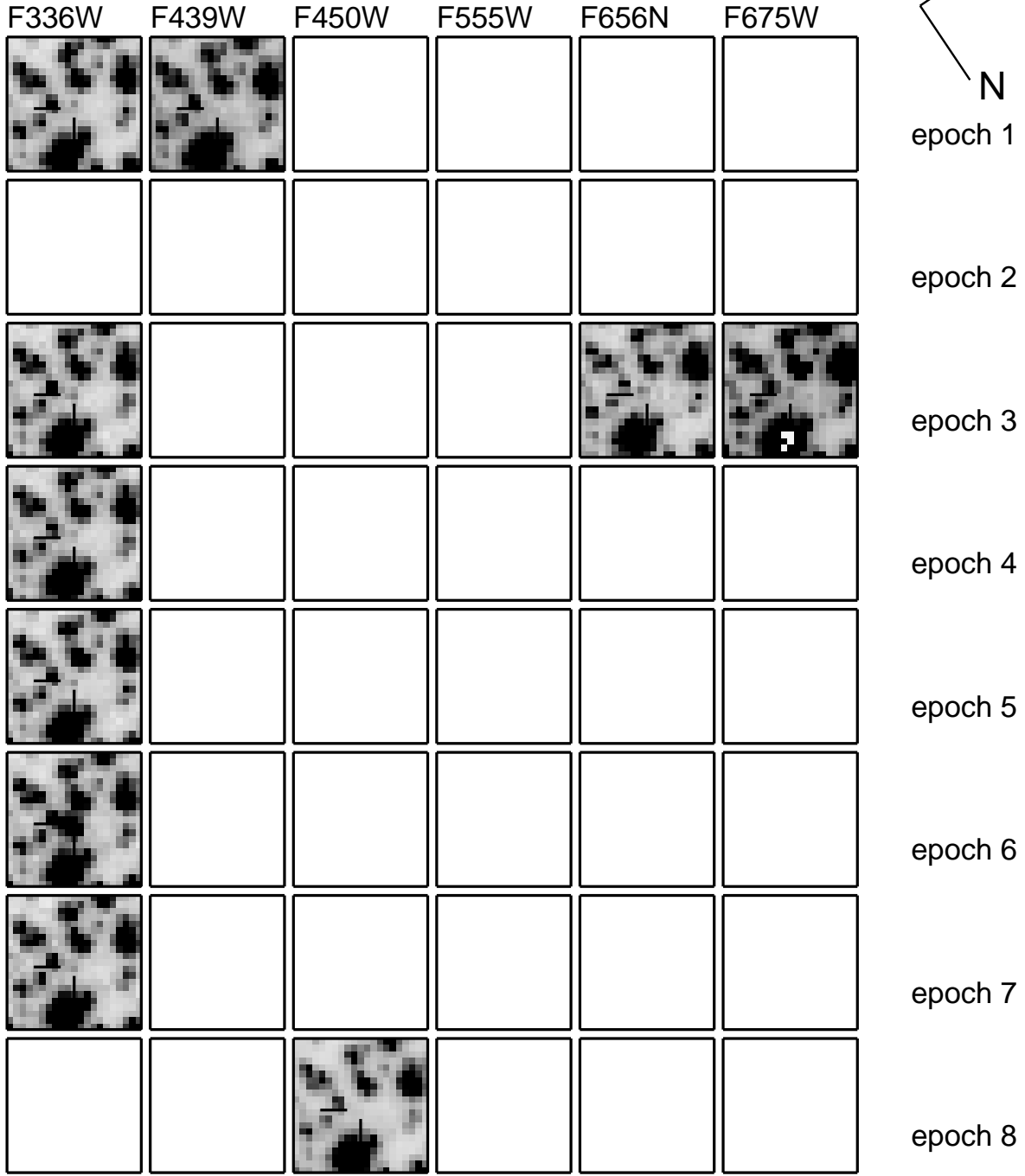}
\caption{Thumbnail images for the second Dwarf Nova (DN2) in M80. It is in eruption in
epoch 6 and just off the PC chip in epoch 2. 
Each thumbnail is 2.1 arcseconds on a side, and North and East are indicated at the upper right. }
\end{figure}

\begin{figure}
\figurenum{8}
\plotone{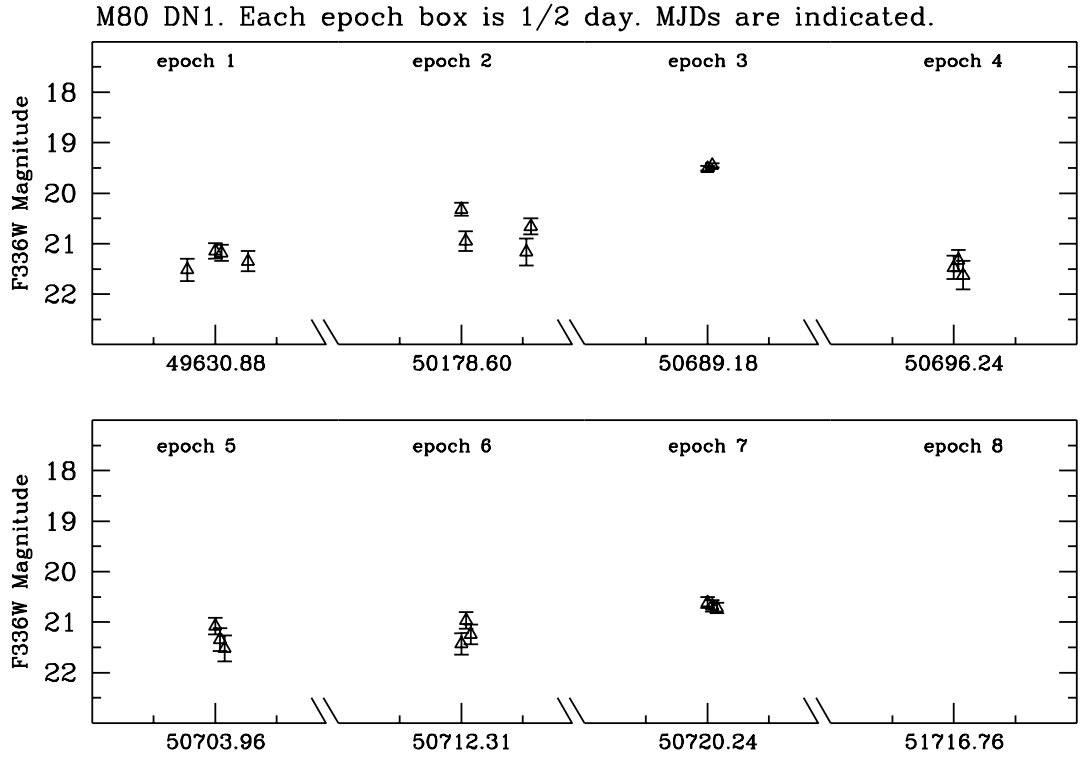}
\caption{The lightcurve for the first Dwarf Nova (DN1) in F336W. The object is in eruption 
in epoch 3.}
\end{figure}

\begin{figure}
\figurenum{9}
\plotone{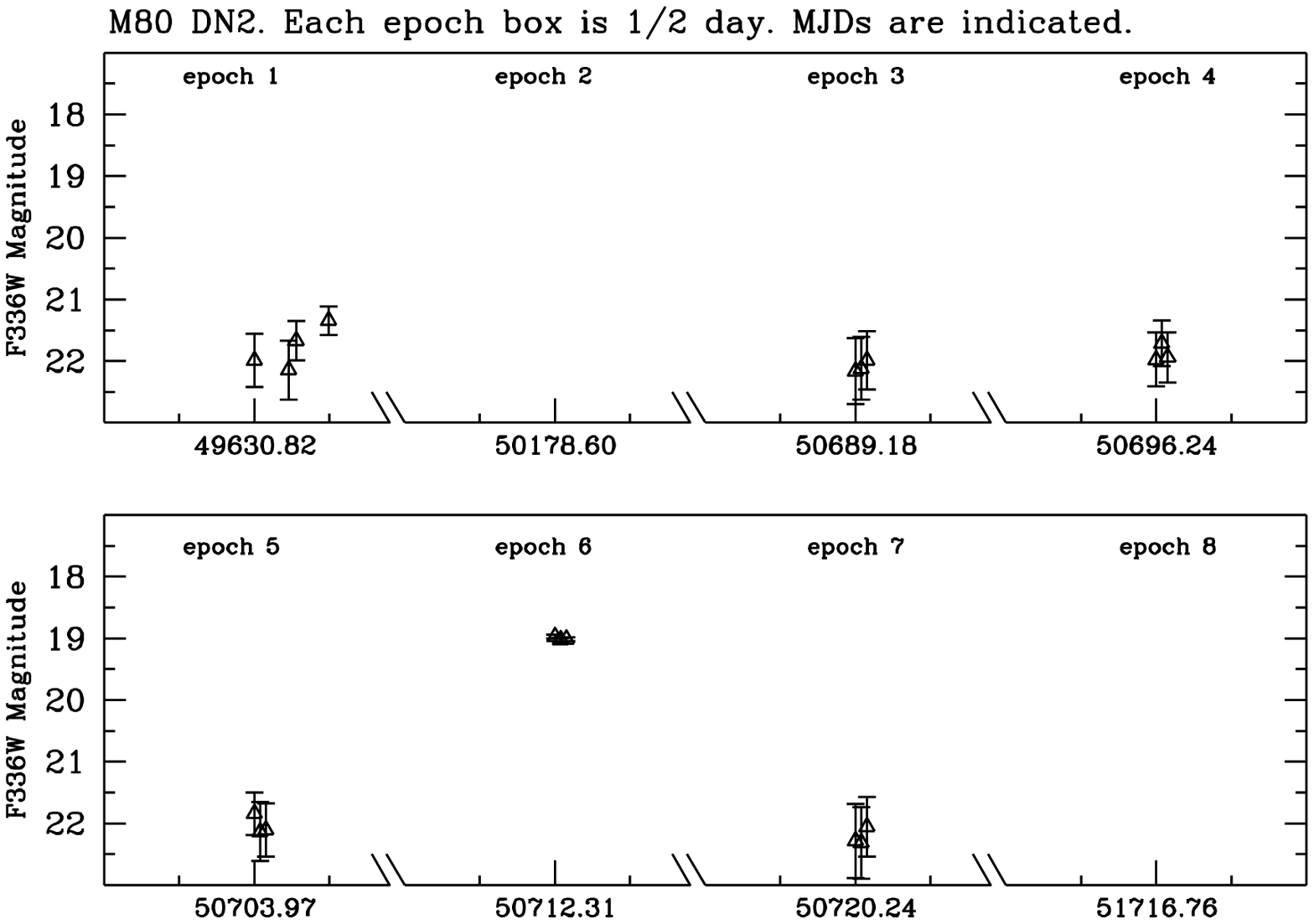}
\caption{The lightcurve for the second Dwarf Nova (DN2) in F336W. The object is in eruption  
in epoch 6, and just outside the field-of-view in epoch 2.}
\end{figure}

\begin{figure}
\figurenum{10}
\plotone{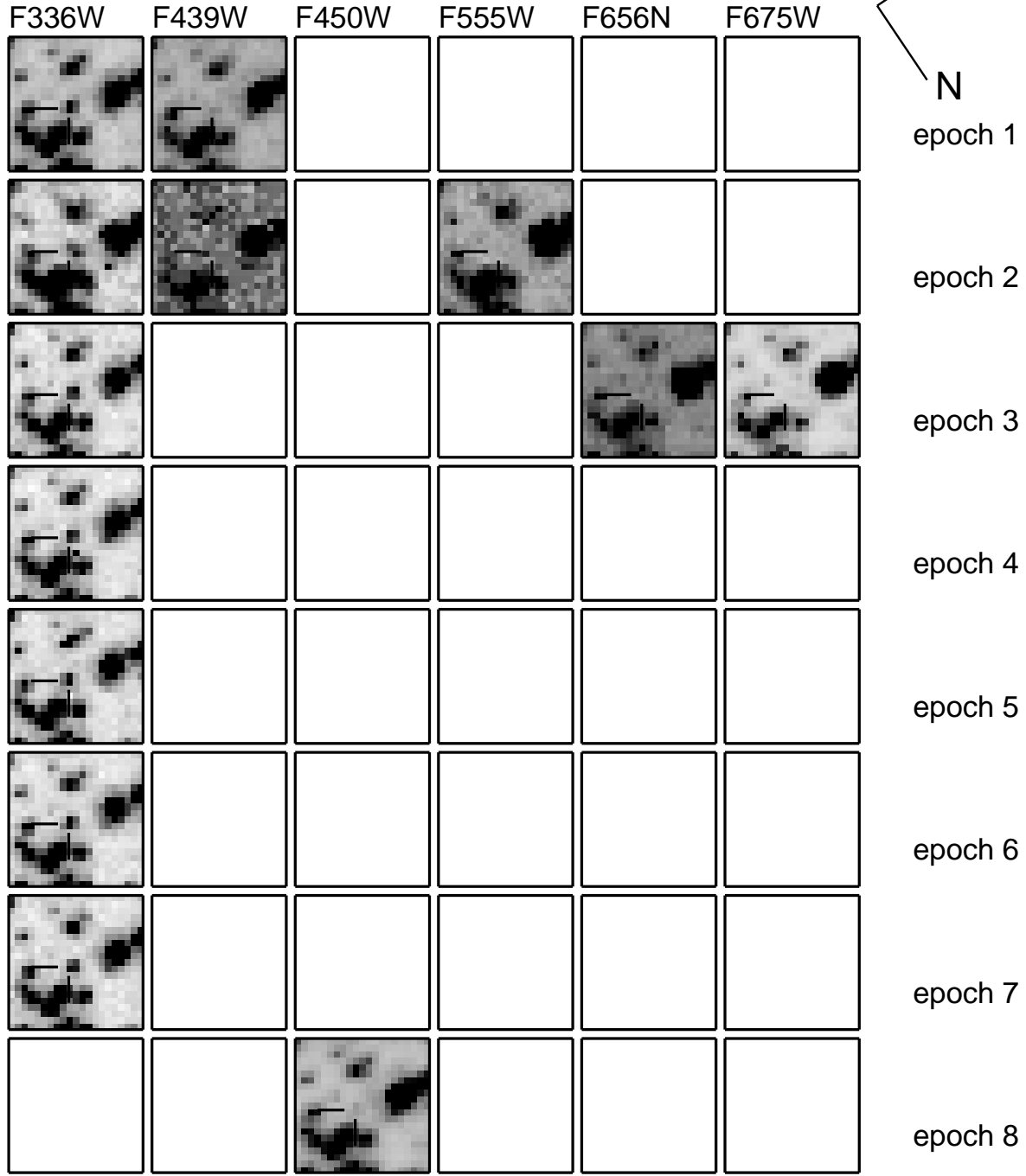}
\caption{Thumbnail images of the classical nova T Sco in M80. These can be compared to Shara and Drissen (1995).
Each thumbnail is 1.04 arcseconds on a side.}
\end{figure}

\begin{figure}
\figurenum{11}
\plotone{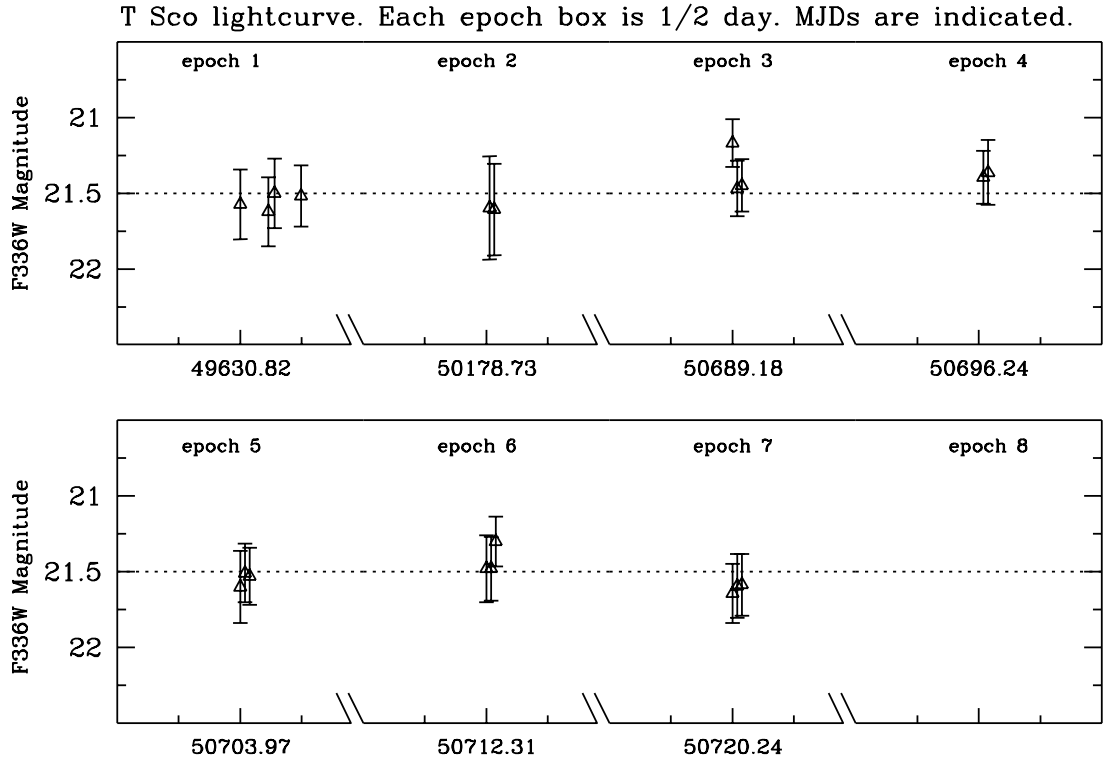}
\caption{F336W Photometry for the old nova T Sco (nova 1860). The object is remarkably constant in brightness. }
\end{figure}

\begin{figure}
\figurenum{12}
\plotone{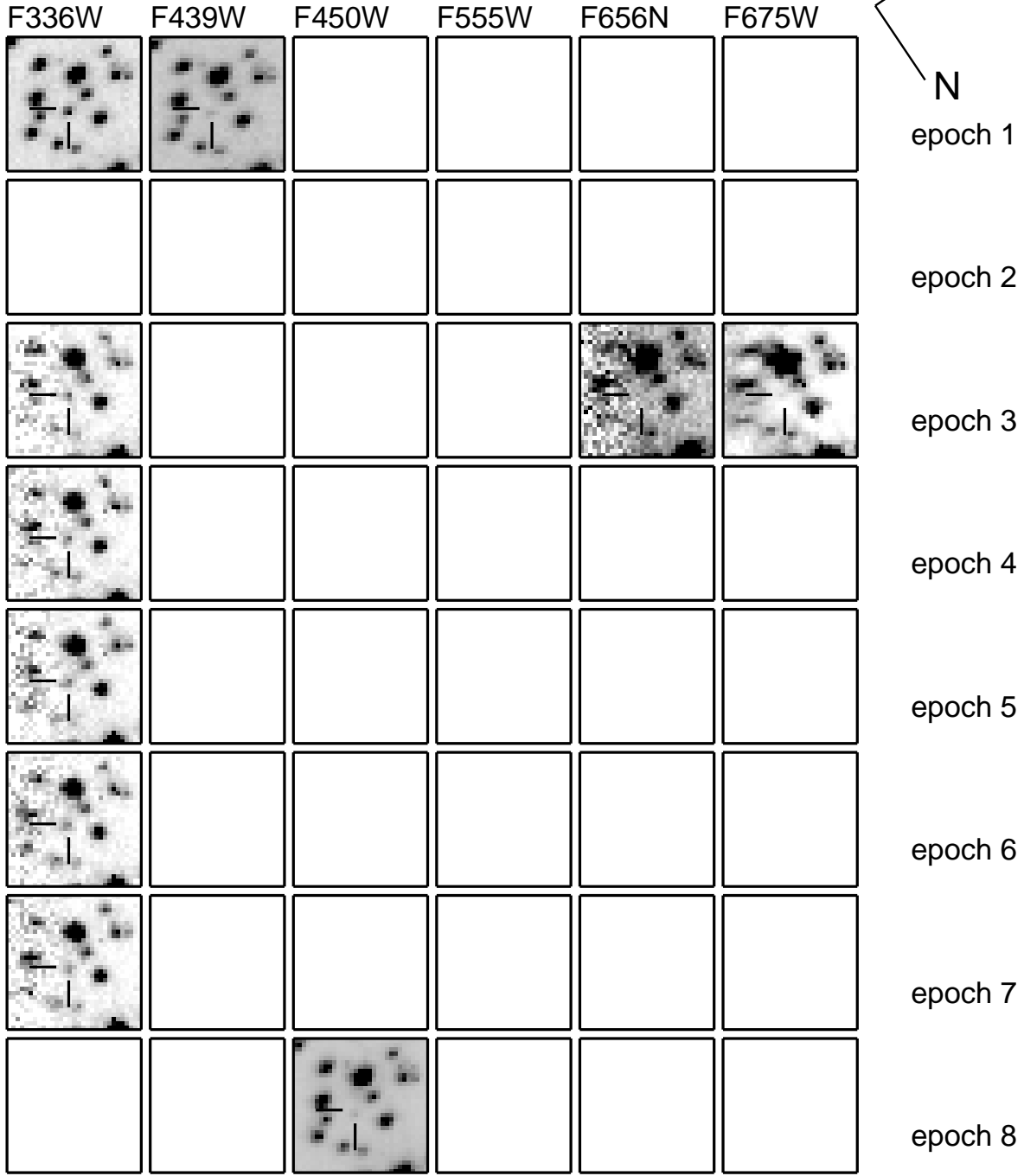}
\caption{Images of the extremely Blue object of \citet{sha95}.
Note that for epochs 3-7, the object is near the edge of the PC chip. The object was just off the chip 
in epoch 2. Each thumbnail is 1.54 arcseconds on a side. }
\end{figure}

\begin{figure}
\figurenum{13}
\plotone{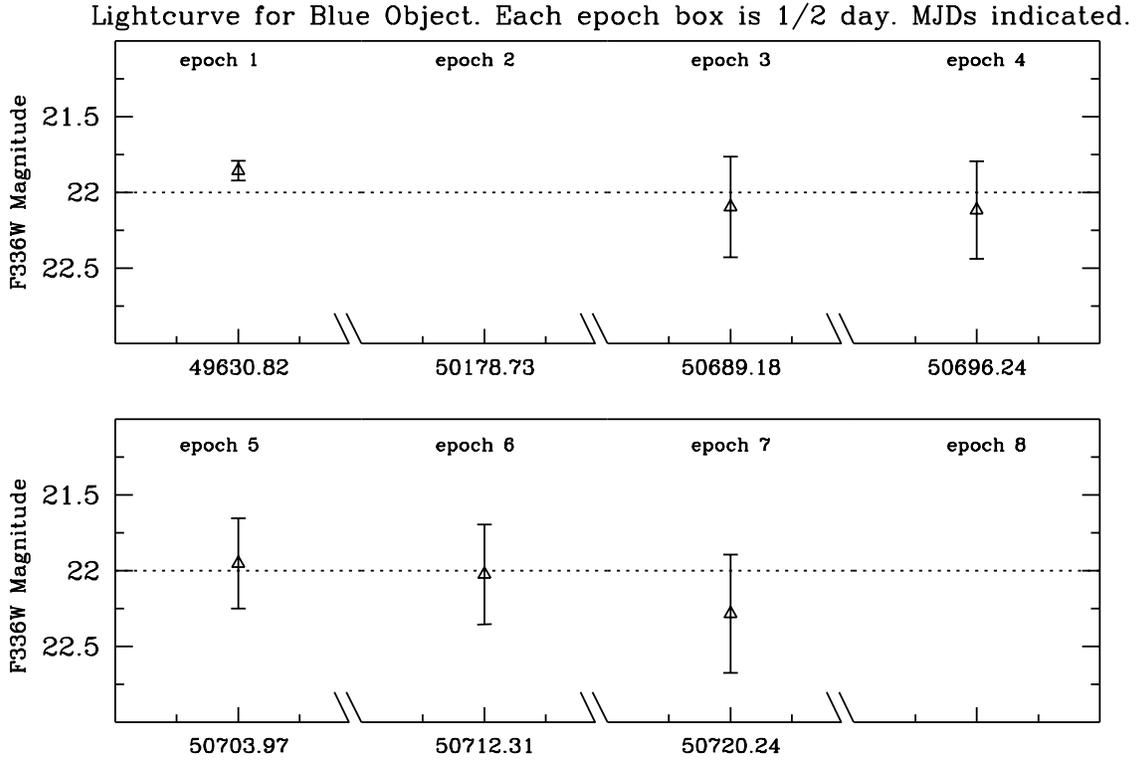}
\caption{An F336W lightcurve for the faint blue object. The large error bars in epochs 3-7 are due to the 
relatively low S/N caused by the object's proximity to the chip edge.}
\end{figure}  

\begin{figure}
\figurenum{14}  
\plotone{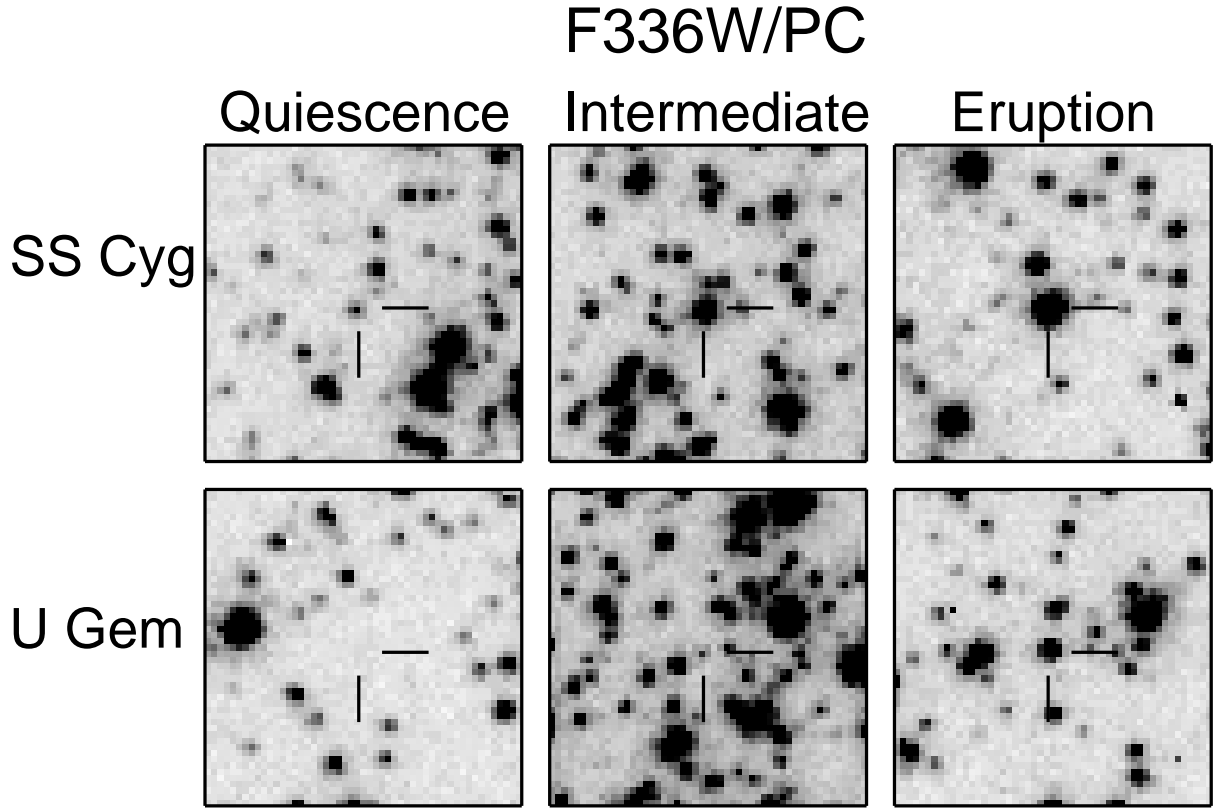}
\caption{Examples of artificial Dwarf Novae in M80. Typical images of the two prototypical Dwarf Novae 
SS Cyg and U Gem, placed in M80 F336W PC images, are shown in three different phases, from 
quiescence to eruption.}
\end{figure}

\begin{figure}
\figurenum{15}
\plotone{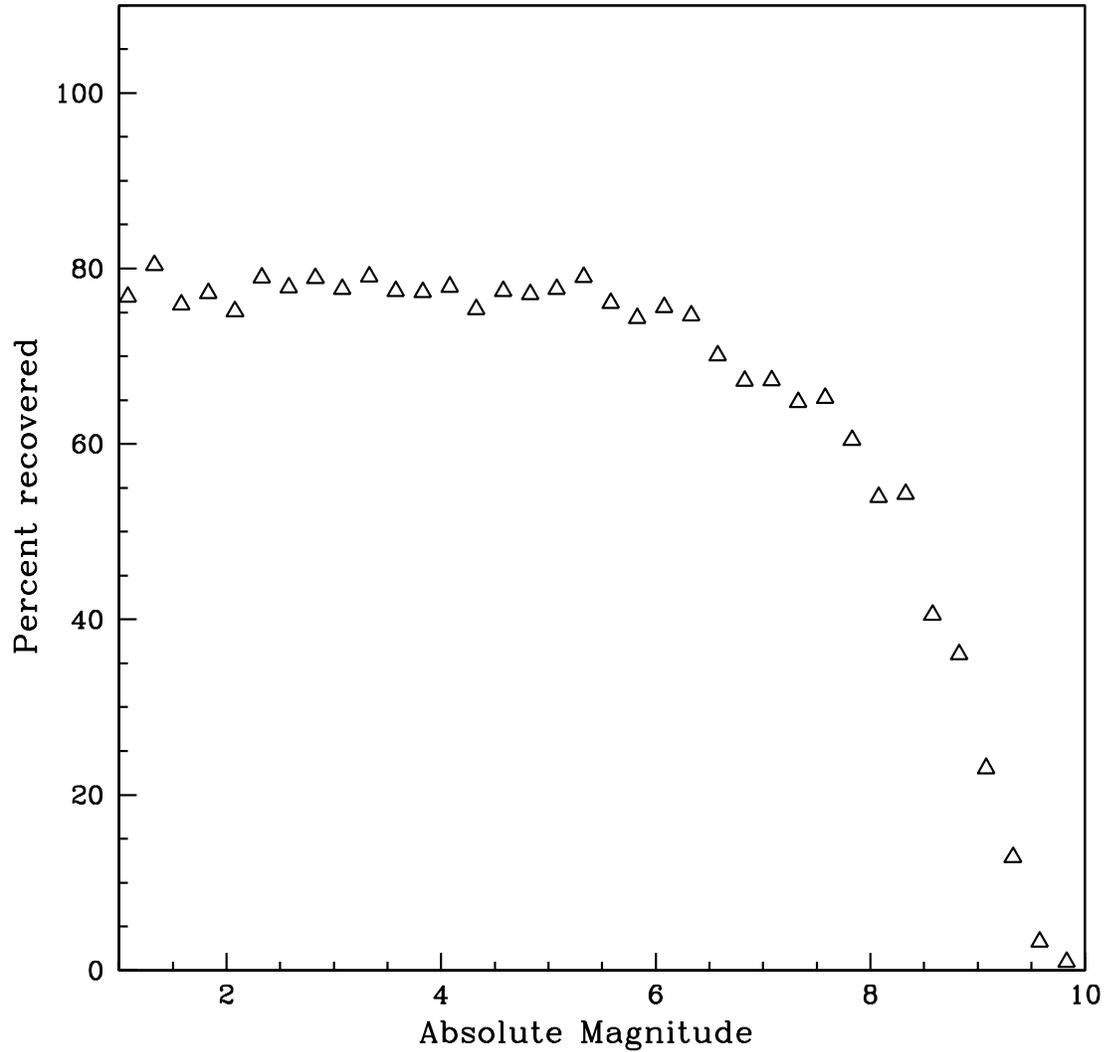}
\caption{Completeness curves for a single-epoch coadded F336W image of M80. The figure assumes an 
M80 distance modulus of $(m - M) = 15.06$. Several hundred artificial stars were added to the 
F336W image in each magnitude bin. Only those stars recovered at the same position AND within 0.5 
mag of the artificial star were considered recovered. Crowding accounts for the $\sim 20\%$ 
incompleteness for magnitudes between 0 and $\sim 6$.} 
\end{figure}

\begin{figure}
\figurenum{16}
\plotone{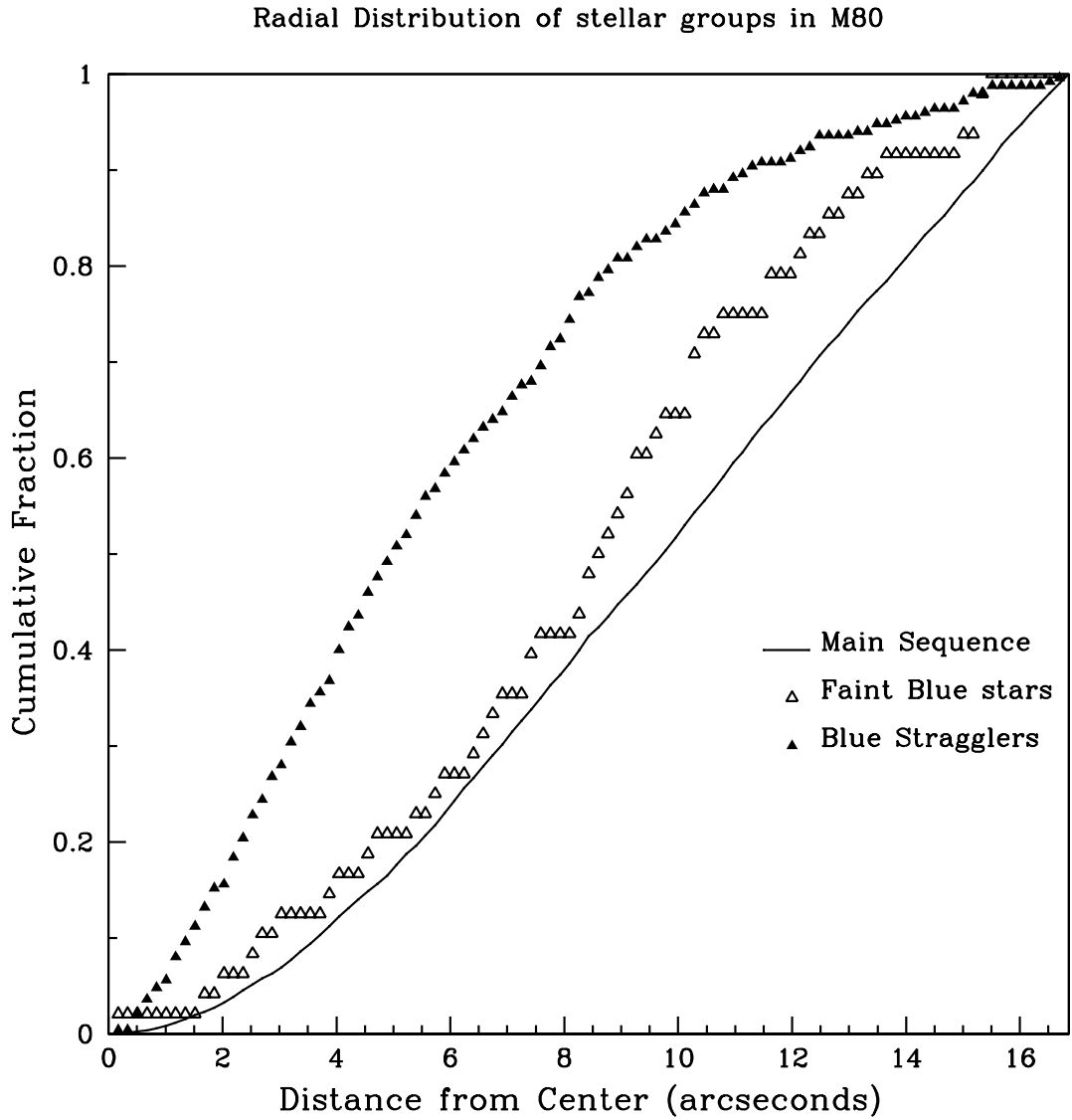}
\caption{The cumulative radial distribution of three different groups of stars as indicated. The 
open triangles correspond to the faint blue stars running nearly parallel to the main sequence
indicated in Figure 5. Their radial distribution differs from that of the main sequence stars 
(from the K-S test) with $>95\%$ likelihood} 
\end{figure}

\begin{figure}
\figurenum{17}
\plotone{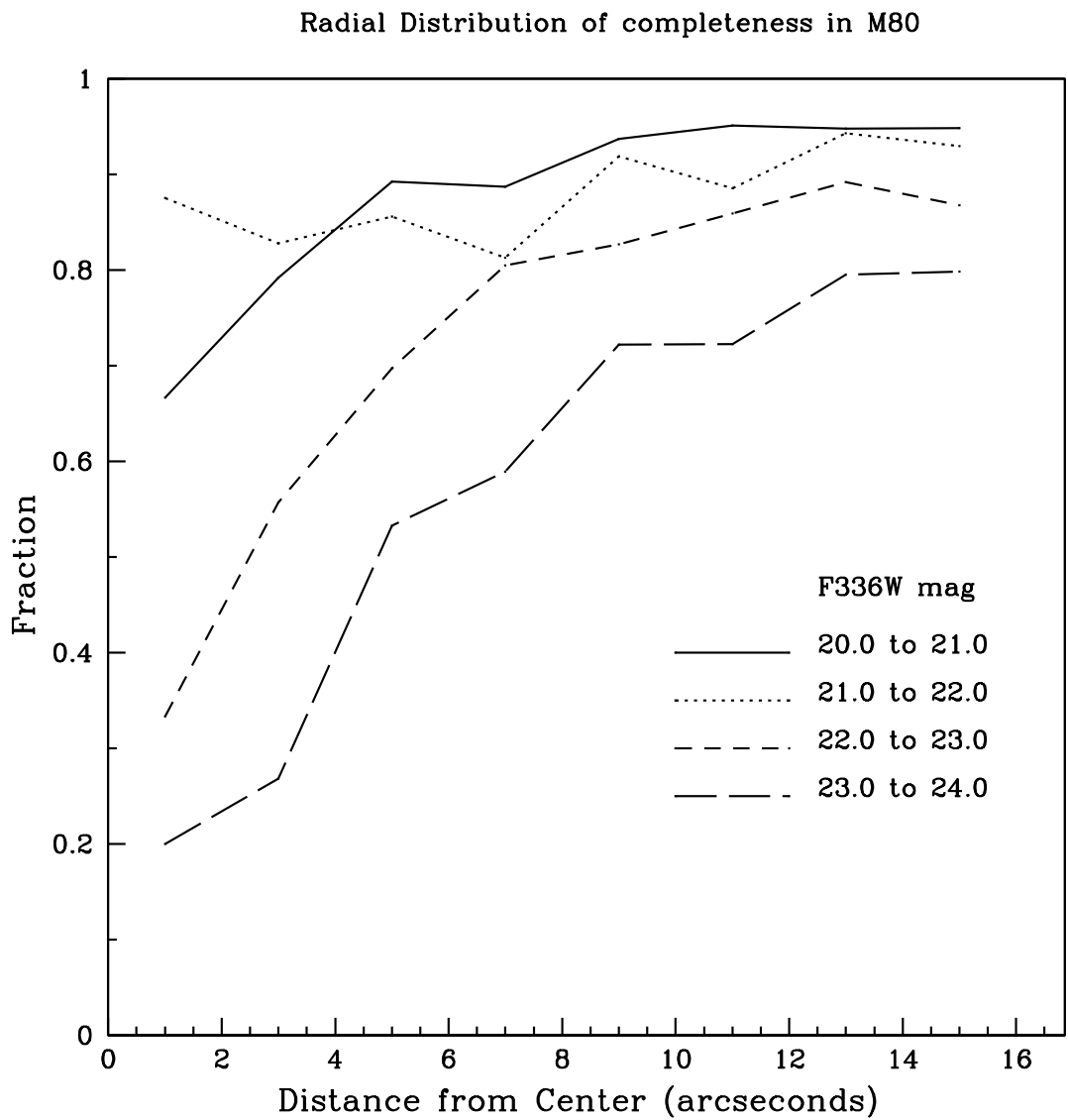}
\caption{A simulation showing the recovery fraction of a sample of artificial stars randomly placed 
on an F336W image as a function of distance from the cluster center. We are significantly incomplete at 
$r < 8$ arcsec for stars fainter than $\simeq 22.0$ magnitude. This further strengthens the suggestion 
(From Figure 16) that the faint blue stars in M80 are centrally concentrated and hence good binary candidates.} 
\end{figure}

\begin{figure}
\figurenum{18}
\plotone{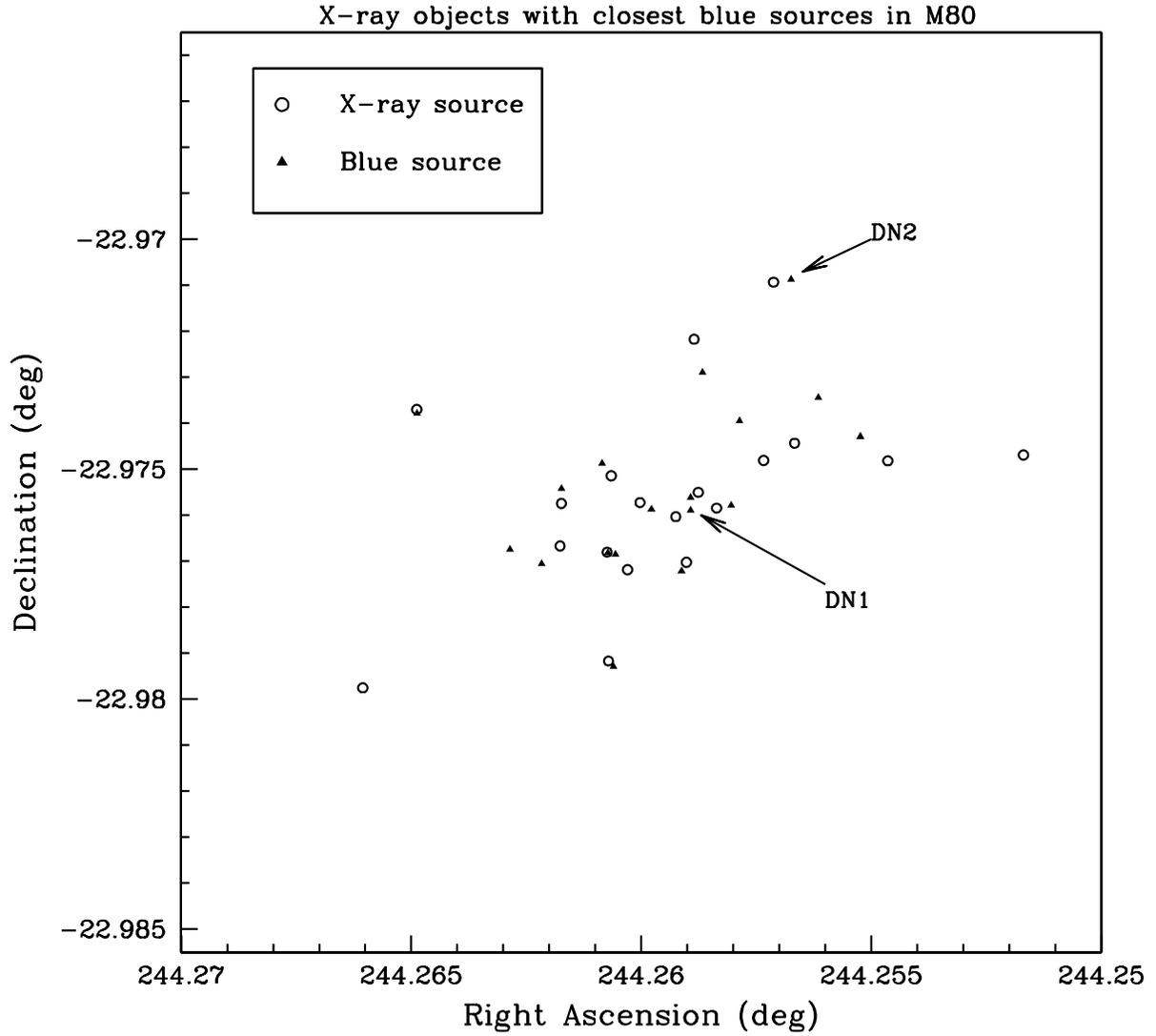}
\caption{A plot showing the positions (circles) of the 19 X-ray bright sources noted by \citet{hei03}. The positions 
(triangles) of the closest faint blue stars are also shown (see text). The locations of Dwarf Nova 1 and 2 are 
marked. } 
\end{figure}

\clearpage

\begin{figure}
\figurenum{19}
\plotone{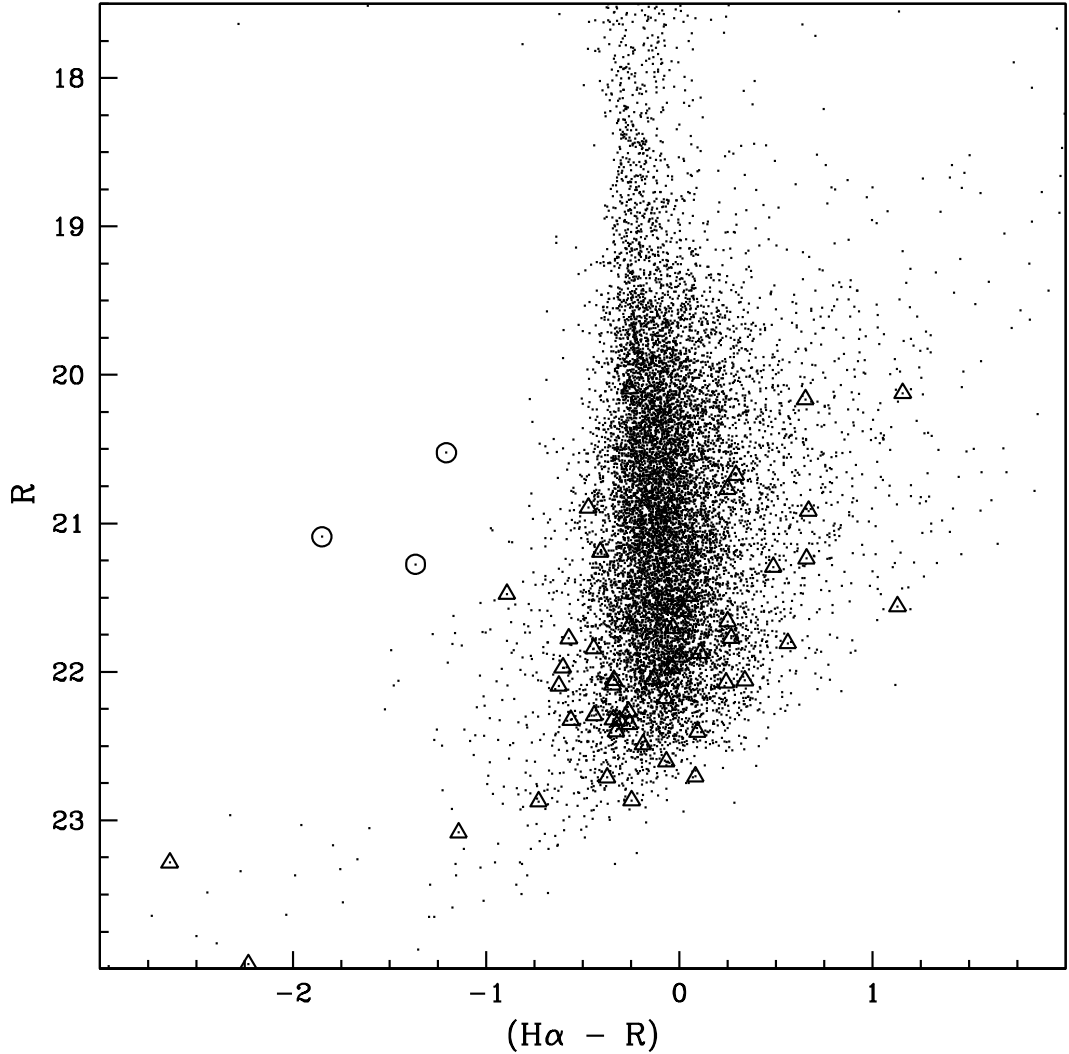}
\caption{An $R$ versus H$\alpha - R$ color-magnitude diagram with 51 of the  faint blue objects (see Fig. 5) 
marked with triangles (see text). None of these faint blue objects appear to have an H$\alpha$ 
luminosity significantly greater than the overall cluster population. The three circled 
objects are heavily crowded stars near the cluster core, and their photometry is thus highly suspect.} 
\end{figure}

\clearpage

\begin{deluxetable}{l|l|c|c|c|c|c|c}
\tabletypesize{\scriptsize}
\tablecaption{Table of M80 observations: Listed are the number of useable observations at each date, as well as the total exposure time.}
\tablewidth{0pt}
\tablehead{\colhead{Epoch} & \colhead{PI/Prog. ID} & \colhead{F336W} & \colhead{F439W} & \colhead{F450W}           & \colhead{F555W}            & \colhead{F656N}    & \colhead{F675W}}
\startdata                            
1: 10/5/94 & Shara  & 4$\times$900s         & 4$\times$300s       &                              &                            &                    &                    \\ 
           & \#5677 &  $\Rightarrow$3600s   &  $\Rightarrow$1200s &                              &                            &                    &                    \\ \hline 
2: 4/5/96  & Ferarro& 4$\times$600s         & 2$\times$30s        &                              & 2$\times$2s, 4$\times$23s  &                    &                    \\ 
           & \#5903 &  $\Rightarrow$2400s   &  $\Rightarrow$60s   &                              &      $\Rightarrow$96s      &                    &                    \\ \hline
3: 8/29/97 & Shara  & 2$\times$700, 800s    &                     &                              &                            &3$\times$1300s      & 3$\times$260s      \\ 
           & \#6460 &  $\Rightarrow$2200s   &                     &                              &                            & $\Rightarrow$3900s & $\Rightarrow$780s  \\ \hline
4: 9/5/97  & Shara  & 2$\times$700, 800s    &                     &                              &                            &                    &                    \\ 
           & \#6460 &  $\Rightarrow$2200s   &                     &                              &                            &                    &                    \\ \hline
5: 9/12/97 & Shara  & 2$\times$700, 800s    &                     &                              &                            &                    &                    \\ 
           & \#6460 &  $\Rightarrow$2200s   &                     &                              &                            &                    &                    \\ \hline
6: 9/21/97 & Shara  & 2$\times$700, 800s    &                     &                              &                            &                    &                    \\ 
           & \#6460 &  $\Rightarrow$2200s   &                     &                              &                            &                    &                    \\ \hline
7: 9/29/97 & Shara  & 2$\times$700, 800s    &                     &                              &                            &                    &                    \\ 
           & \#6460 &  $\Rightarrow$2200s   &                     &                              &                            &                    &                    \\ \hline
8: 6/21/00 & King   &                       &                     &5$\times$160s, 7$\times$ 200s &                            &                    &                    \\ 
           & \#8655 &                       &                     &    $\Rightarrow$2200s        &                            &                    &                    \\ 
\enddata
\end{deluxetable}

\clearpage

\begin{deluxetable}{c|ll||ll|r|c}
\tabletypesize{\scriptsize}
\tablecaption{List of the 19 bright X-ray sources from Heinke et al. (2003) with the closest blue objects, and
their separation in arcseconds, after applying the offsets discussed in the text. }
\tablewidth{0pt}
\tablehead{
\colhead{Source} & \colhead{R.A. (X-ray)} & \colhead{Decl. (X-ray)} & \colhead{R.A. (blue)} & 
\colhead{Decl. (blue)} & \colhead{$\Delta$} & \colhead{Notes}}
\startdata
CX1  & 16 17 02.814  &  -22 58 32.67  &  16 17 2.813  &   -22 58 31.51 &  1.160    &    \\ 
CX2  & 16 17 02.576  &  -22 58 36.48  &  16 17 2.573  &   -22 58 36.55 &  0.085    &    \\ 
CX3  & 16 17 01.597  &  -22 58 27.95  &  16 17 1.474  &   -22 58 24.38 &  3.969    &    \\ 
CX4  & 16 17 02.005  &  -22 58 33.03  &  16 17 1.928  &   -22 58 32.80 &  1.086    &    \\ 
CX5  & 16 17 01.708  &  -22 58 15.34  &  16 17 1.612  &   -22 58 15.13 &  1.282    & DN2\\ 
CX6  & 16 17 03.569  &  -22 58 25.30  &  16 17 3.569  &   -22 58 25.61 &  0.302    &    \\ 
CX7  & 16 17 02.164  &  -22 58 37.27  &  16 17 2.189  &   -22 58 37.95 &  0.766    &    \\ 
CX8  & 16 17 01.114  &  -22 58 29.33  &  16 17 1.255  &   -22 58 27.44 &  2.731    &    \\
CX9  & 16 17 02.401  &  -22 58 32.60  &  16 17 2.346  &   -22 58 33.13 &  0.967    &    \\ 
CX10 & 16 17 00.407  &  -22 58 28.87  &  16 17 1.255  &   -22 58 27.44 & 11.820    &    \\ 
CX11 & 16 17 02.472  &  -22 58 37.86  &  16 17 2.532  &   -22 58 36.66 &  1.483    &    \\ 
CX12 & 16 17 02.565  &  -22 58 45.00  &  16 17 2.539  &   -22 58 45.44 &  0.564    &    \\ 
CX13 & 16 17 01.755  &  -22 58 29.29  &  16 17 1.880  &   -22 58 26.22 &  3.524    &    \\ 
CX14 & 16 17 02.553  &  -22 58 30.50  &  16 17 2.601  &   -22 58 29.53 &  1.176    &    \\ 
CX15 & 16 17 02.100  &  -22 58 31.80  &  16 17 2.141  &   -22 58 32.19 &  0.680    &    \\ 
CX16 & 16 17 02.119  &  -22 58 19.80  &  16 17 2.079  &   -22 58 22.40 &  2.666    &    \\ 
CX17 & 16 17 02.220  &  -22 58 33.70  &  16 17 2.141  &   -22 58 33.20 &  1.174    & DN1\\ 
CX18 & 16 17 02.820  &  -22 58 36.00  &  16 17 2.916  &   -22 58 37.38 &  1.901    &    \\ 
CX19 & 16 17 03.850  &  -22 58 47.10  &  16 17 3.081  &   -22 58 36.26 & 15.172    &    \\ 
\enddata
\end{deluxetable}

\end{document}